\documentclass[12pt]{iopart}

\usepackage{iopams}
\usepackage{setspace}

\usepackage{amsfonts}   % note how statements can be commented out
\usepackage{amssymb}
\usepackage{graphicx,epstopdf}   % for figures

\begin{document}

\title{An exactly solvable toy model}

\author{Wang X G$^{1, 2, 3}$ and Zhang J M$^{1,2}$}

\address{$^1$ Fujian Provincial Key Laboratory of Quantum Manipulation and New Energy Materials,
College of Physics and Energy, Fujian Normal University, Fuzhou 350007, China}
\address{$^2$ Fujian Provincial Collaborative Innovation Center for Optoelectronic Semiconductors and Efficient Devices, Xiamen 361005, China}
\address{$^3$ Institute of Physics, Chinese Academy of Sciences, Beijing 100080, China}

\begin{abstract}
In an attempt to regularize a previously known exactly solvable model [Yang and Zhang, Eur. J. Phys. \textbf{40}, 035401 (2019)], we find yet another exactly solvable toy model. The interesting point is that while  the Hamiltonian of the model is parameterized by a function $f(x)$ defined on $[0, \infty )$, its spectrum depends only on the end values of $f$, i.e., $f(0)$ and $f(\infty )$. This model can serve as a good exercise in quantum mechanics at the undergraduate level.
\end{abstract}

%\submitto{\EJP}
\maketitle

\section{Introduction}

Recently, in studying the quench dynamics of a Bloch state in a one-dimensional tight-binding ring \cite{epl1,epl2,prb}, an exactly solvable model was devised and its dynamics was solved. Later, the eigenstates and eigenenergies of the model were solved in closed and neat forms \cite{ejp19}.

The model is very simple. It consists of infinitely many levels $\{|n \rangle , n\in \mathbb{Z} \}$, with the Hamiltonian being
\begin{eqnarray}\label{H0}
% \nonumber to remove numbering (before each equation)
  H  = \sum_{n = -\infty}^\infty n \Delta |n \rangle \langle n | +  g \sum_{n_1,n_2 = -\infty}^\infty  |n_1 \rangle \langle n_2 | .
\end{eqnarray}
The first term is diagonal. It simply means that the levels are equally spaced with the spacing being $\Delta $. The second term is off-diagonal. Its peculiarity is that it couples two arbitrary levels, regardless of their distance in energy, with a constant strength $g$. This makes it a rank-1 operator \footnote{In linear algebra, a $p\times  q $ rank-1 matrix is the product of a $p$-component column vector and a $q$-component row vector. Here, the off-diagonal term can obviously be written as $g |u \rangle \langle u | $, with $|u\rangle = \sum_{n\in \mathbb{Z}} |n \rangle $. }, a  fact which was believed to be responsible for the exact solvability of the model.

It should be instructive to review how this ideal, contrived model (\ref{H0}) arises in a more realistic problem. In \cite{epl1,epl2,prb}, we studied such a dynamical problem. The setting is a tight-binding lattice \cite{mermin} with the Hamiltonian
\begin{equation}\label{htb}
  H_{tb} = -\sum_{m=0}^{L -1 } (|m\rangle \langle m+1 | + |m+1\rangle \langle m |).
\end{equation}
Here $\{|m\rangle  \}$ are a set of orthonormal, localized single-particle orbits. The Hamiltonian simply means that the particle can hop between two adjacent sites. We take the periodic boundary condition so that $|0\rangle = |L \rangle $. Equivalently, one can imagine that the sites are arranged into a ring. The eigenstates of $H_{tb}$ are simply plane waves (Bloch states) on the ring,
\begin{equation}\label{blochstate}
  |\overline{k}\rangle = \sum_{m=0}^{L-1} |m \rangle \langle m| \overline{k}\rangle = \frac{1}{\sqrt{L}} \sum_{m=0}^{L-1} e^{i 2 \pi k m/L} |m \rangle ,
\end{equation}
where $k$ is an integer. Apparently, $|\overline{k}\rangle = | \overline{k + L} \rangle $. The corresponding eigenenergy is $\varepsilon (k) = -2 \cos (2\pi k/L)$.

Now consider such a problem. Suppose initially the particle is in a generic Bloch state, i.e., $|\phi(t=0)\rangle = |\overline{k_i}\rangle $, where $k_i $ is an arbitrary integer. It is stationary with respect to $H_{tb}$. However, at $t=0$, we quench it by adding a potential to an arbitrary site of the ring. Without of loss of generality (since all the sites are on an equal footing), let this site be $m=0$. The Hamiltonian is now $H_{tb} + H_{pot}$, with
\begin{equation}\label{hpot}
  H_{pot} = V |0 \rangle \langle 0| ,
\end{equation}
where $V$ is the potential strength. The initial state is no longer an eigenstate of the post-quench Hamiltonian  and it starts evolving. Many interesting phenomena ensue.  For example, it is found that the survival probability of the initial state, namely $|\langle \phi(t=0)| \phi(t) \rangle |^2$, display cusps periodically in time.

To account for these phenomena, we notice two salient features of the problem. First, it is observed numerically that at any time, only those Bloch states with energy close to that of the initial Bloch state are significantly populated. That is, only those Bloch states $|\overline{k} \rangle $ with $k\simeq k_i$ or $k\simeq -k_i$ are actively participating in the dynamics. These states are almost equally spaced in energy, with the gap being $( \partial \varepsilon /\partial k )|_{k=k_i}= (4 \pi /L) \sin (2 \pi k_i/L)$. Second, as can be easily verified, the quench potential $H_{pot}$, which induces the nontrivial dynamics, couples two arbitrary Bloch states with a constant strength. Specifically,
\begin{eqnarray}\label{hpotcouple}
% \nonumber to remove numbering (before each equation)
  \langle \overline{k_1} |H_{pot} | \overline{k_2} \rangle  &=& \sum_{m=0}^{L-1} \langle \overline{k_1}| m \rangle \langle m  |H_{pot} |m\rangle \langle m | \overline{k_2} \rangle \nonumber \\
  &=& \frac{1}{L } \sum_{m=0}^{L-1}\langle m  |H_{pot} |m\rangle e^{i 2 \pi ({k_2} - {k_1}) m /L  } = \frac{V}{L},
\end{eqnarray}
\emph{regardless of} the values of $k_{1,2}$.

It is these two features that motivated the construction of the model (\ref{H0}), which successfully accounted for all the dynamical phenomena observed in the original tight-binding model.

\section{A more regular model}

From the above, specifically (\ref{hpotcouple}), we see that the constant coupling comes from a very special potential, which is supported on a single site. For a generic, more extended potential, the coupling cannot be a constant. Actually, from (\ref{hpotcouple}), we see that the coupling strength between two Bloch states is just the Fourier transform of the quench potential. Now by the Riemann-Lebesgue lemma \cite{bocher}, if the quench potential is more extended, its Fourier transform would decay to zero in the high frequency limit. We are thus motivated to ``regularize'' the off-diagonal coupling term in (\ref{H0}) and consider the following Hamiltonian
\begin{eqnarray}\label{H}
% \nonumber to remove numbering (before each equation)
  H_f  = \sum_{n = -\infty}^\infty n \Delta |n \rangle \langle n | +  \sum_{n_1,n_2 = -\infty}^\infty f(n_1 - n_2) |n_1 \rangle \langle n_2 | ,
\end{eqnarray}
where $f(x)$ is a real, even function defined on the real line satisfying the condition $\lim_{x\rightarrow \infty} f(x) = 0 $.

The coupling term is now no longer a rank-1 perturbation, and the approach in \cite{ejp19} fails. But it is still regular. Actually, it is in the form of a Toeplitz matrix \footnote{A Toeplitz matrix, or a diagonal-constant matrix, is a structured  matrix in which each descending diagonal from left to right is constant. See B\"ottcher A and Grudsky S M 2012 Toeplitz Matrices, Asymptotic Linear Algebra, and Functional Analysis (Birkh\"auser, Basel).} and its action on a wave function is to convolve it with the function $f$. This observation suggests the Fourier transform\footnote{As a well-known theorem, the Fourier transform of the convolution of two functions is the pointwise product of their Fourier transforms. That is, two functions convolved in the real space are factorized in the momentum space. This is of course a great simplification. },  and we solve the eigenvalues as
\begin{eqnarray}\label{spectrum1}
% \nonumber to remove numbering (before each equation)
  \varepsilon_m  &=&  m \Delta  +f(0) ,  \quad m \in \mathbb{Z} .
\end{eqnarray}
Here it is interesting that while the Hamiltonian (\ref{H}) is defined with the function $f$ as a parameter as a whole, its spectrum depends only on the end value $f(0)$.

The Fourier transform approach works also for the original Hamiltonian (\ref{H0}). We thus find that for the function $f$ we can drop the condition that it converges to zero as $x\rightarrow \pm \infty $, and require just that the limit exists. In this case, the spectrum is
\begin{eqnarray}\label{spectrum2}
% \nonumber to remove numbering (before each equation)
  \varepsilon_m  &=& m \Delta +f(0)- f(\infty) + \frac{\Delta }{\pi }\arctan \left( \frac{\pi f(\infty )}{\Delta } \right) , \quad m \in \mathbb{Z} .
\end{eqnarray}
It depends only on the end values $f(0)$ and $f(\infty )$. For the original Hamiltonian (\ref{H0}), $f(0) =f(\infty ) =g$, and we recover the results in \cite{ejp19}.

\section{Analytical solution of the eigenvalue problem}

%We now proceed to solve the eigenvalues of $H_\lambda$.

First of all, as in \cite{ejp19}, we notice that the off-diagonal part is invariant under the translation $|n \rangle \rightarrow |n + 1 \rangle $, which implies that the energy spectrum should be equally spaced with the equal gap being $\Delta $. Formally, let us define the translation or raising operator $   T =   \sum_{n = -\infty}^\infty  |n+1 \rangle \langle n | $. We claim that if $|\psi\rangle $ is an eigenstate of $H_f$ with eigenvalue $\varepsilon $, then $T | \psi \rangle $ is an eigenstate with eigenvalue $\varepsilon + \Delta $.\footnote{At this point, one might jump to the conclusion that the spectrum of the model must be in the form of (\ref{spectrum1}), as we have $Tr(H_f)=\sum_{m\in \mathbb{Z}} [ m \Delta + f(0)] =  \sum_{m\in \mathbb{Z}} \varepsilon_m $. But this identity is derived for a finite matrix. A mindless generalization to the infinite case is dangerous. Actually, it is even a nontrivial problem to make sense of the trace of an infinite matrix. } This is because we have the commutation relation $  [H_f, T] = \Delta  T $ and hence
\begin{eqnarray}
% \nonumber to remove numbering (before each equation)
  H_f T |\psi\rangle  &=& (T H_f +  \Delta T ) |\psi \rangle = (T  \varepsilon  +  \Delta T ) |\psi \rangle = (\varepsilon  + \Delta  )T |\psi \rangle.
\end{eqnarray}

To determine the spectrum, we have to consider the Schr\"odinger equation $H_f |\psi \rangle = \varepsilon  | \psi \rangle $. Assuming that  $|\psi \rangle = \sum_{n} |n \rangle \langle n |\psi \rangle =  \sum_n a_n  | n \rangle $, we have
\begin{eqnarray}\label{seq1}
% \nonumber to remove numbering (before each equation)
  \varepsilon  a_n  = n \Delta  a_n +   \sum_{m =-\infty}^\infty f( n- m )a_m .
\end{eqnarray}
The second term on the right hand side is in the form of (discrete) convolution. This suggests the Fourier transform,
\begin{eqnarray}
% \nonumber to remove numbering (before each equation)
  A_k  = \sum_{n = -\infty}^\infty e^{ikn } a_n ,\quad F_k = \sum_{n = -\infty}^\infty e^{ikn } f(n).
\end{eqnarray}
For the left hand side of (\ref{seq1}), we get $\varepsilon A_k$. For the right hand side, we get
\begin{eqnarray}
% \nonumber to remove numbering (before each equation)
  & & \sum_{n =-\infty}^\infty e^{ikn } \Delta  n  a_n + \sum_{n =-\infty}^\infty  \sum_{m =-\infty}^\infty  e^{ikn }f( n- m ) a_m \nonumber \\
   &=& -i \Delta \frac{\partial}{\partial k } \sum_{n =-\infty}^\infty e^{ikn }  a_n + \sum_{n -m=-\infty}^\infty   e^{ik(n-m) }f( n- m )  \sum_{m =-\infty}^\infty e^{ikm } a_m \nonumber \\
   &=& -i \Delta \frac{\partial}{\partial k }A_k + F_k A_k .
\end{eqnarray}
We thus get the equation of $A_k$,
\begin{eqnarray}\label{eqAk}
% \nonumber to remove numbering (before each equation)
  \varepsilon A_k  &=&  -i\Delta \frac{\partial}{\partial k }A_k +  F_k A_k .
\end{eqnarray}
Note that this is a Dirac-type equation, in the sense that the Hamiltonian on the right hand side is linear in the momentum.

Let us first consider the special case of $f(\infty ) = 0$. In this case, as long as $f$ decays sufficiently fast, $F$ is a regular function of $k$, and we can solve $A$ as
\begin{eqnarray}\label{solFk}
% \nonumber to remove numbering (before each equation)
  A_k  &=& A_0 \exp\left[ \frac{i}{\Delta } \int_0^k dq \left( \varepsilon -F_q \right ) \right] .
\end{eqnarray}
The value of $\varepsilon $ is determined by the boundary condition of $A_0 = A_{2\pi}$, which is equivalent to
\begin{eqnarray}
% \nonumber to remove numbering (before each equation)
  2\pi m \Delta   &=& 2 \pi \varepsilon - \int_0^{2\pi } F_k d k
\end{eqnarray}
for some integer $m$. But the integral here is simply $2\pi f(0)$. We thus obtain the eigenvalues as in (\ref{spectrum1}).
They are indeed equidistant. But what is interesting is that there is only dependence on the end value  $f(0)$.

In the more general case of $f(\infty )\neq 0 $, we can decompose the function into two parts as $f = [f- f(\infty ) ]+ f(\infty )$. The Fourier transform of $f$ is then in the form of
\begin{eqnarray}\label{Fkkk}
% \nonumber to remove numbering (before each equation)
  F_k &=& F^{(1)}_k + 2 \pi f(\infty )\sum_{m \in \mathbb{Z}} \delta(k - 2 \pi m ) ,
\end{eqnarray}
where the first term comes from $f- f(\infty)$, i.e., $F^{(1)}(k) =\sum_{n \in \mathbb{Z}} e^{ikn } [f(n)- f(\infty)] $, while  the delta functions  from the constant part. We still have (\ref{eqAk}). But now $A$ is discontinuous at $k=0$. Integrating from $0^-$ to $0^+$, we get
\begin{eqnarray}\label{link1}
% \nonumber to remove numbering (before each equation)
  0 &=& -i\Delta (A_{0^+} -A_{0^-}) + 2\pi f(\infty ) \int_{0^-}^{0^+}  \delta(k) A(k)  dk \nonumber \\
  &=&   -i \Delta (A_{0^+} -A_{0^-}) + \pi f(\infty ) (A_{0^+} + A_{0^-}),
\end{eqnarray}
i.e.,
\begin{eqnarray}\label{link3}
% \nonumber to remove numbering (before each equation)
  A_{0^+} &=& \frac{\Delta -i \pi f(\infty)}{\Delta + i \pi f(\infty ) }A_{0^-}= \exp \left\{ -2 i \arctan \frac{\pi f(\infty )}{\Delta }  \right \}A_{0^-} .
\end{eqnarray}
On the other hand, integrating from $0^+$ to $2\pi^-$, we get
\begin{eqnarray}\label{link2}
% \nonumber to remove numbering (before each equation)
   A_{0^-} = A_{2\pi^-} &=& A_{0^+}  \exp\left[\frac{ i}{\Delta } \int_0^{2\pi} dk \left( \varepsilon -F_k^{(1)} \right ) \right] \nonumber \\
  &=&A_{0^+} \exp \left[\frac{ i}{\Delta }2\pi (\varepsilon - f(0) + f(\infty )) \right] .
\end{eqnarray}
From (\ref{link3}) and (\ref{link2}), we solve the spectrum in (\ref{spectrum2}). Note that we have rederived the spectrum of the Hamiltonian (\ref{H0}) by exploiting its Toeplitz property rather than its rank-1 property.

Here some mathematical subtlety is worth commenting on. The integral in (\ref{link1}) is actually not well defined. The $\delta$ function, as a distribution, is supposed to act on a smooth function, yet $A$ is discontinuous at $k=0$ as we see in (\ref{link3}) \emph{a posterior}. To arrive at the second line, we defined the product between the delta function and the Heaviside function,  $\delta(k) \theta(k) $, to be $ \frac{1}{2}\delta(k)$. This seems reasonable, as we can ``derive'' it as $\delta \theta = \theta' \theta = \frac{1}{2} (\theta^2)' = \frac{1}{2} \theta' = \frac{1}{2}\delta$. But a second thought would lead also to $\delta \theta = \theta'\theta =  \theta' \theta^n = \frac{1}{n+1} (\theta^{n+1})'= \frac{1}{n+1} \theta' =  \frac{1}{n+1}\delta $ for any positive integer $n$. This ambiguity shows that the product $\delta \theta $ is not well defined as a distribution and the treatment in (\ref{link1}) is questionable.
A different approach is to consider the delta function in (\ref{Fkkk}) as the limit of some ordinary function. Instead of (\ref{link3}), this would lead to the linking condition
\begin{eqnarray}\label{link4}
% \nonumber to remove numbering (before each equation)
  A_{0^+} &=& \exp \left[ \frac{i}{\Delta}\int_{0^-}^{0^+} dq (\varepsilon - F_q)  \right ] A_{0^-} = \exp\left[ -\frac{2\pi i f(\infty )}{\Delta}  \right ] A_{0^-}  ,
\end{eqnarray}
and eventually result in the spectrum of (\ref{spectrum1}) rather than that of (\ref{spectrum2}).

We note that in the literature dealing with the relativistic Dirac equation \cite{glushkovbook}, both approaches exist. In \cite{jpc,surf}, the first approach was taken, which was criticized in \cite{prc}, where the second approach was favored. But actually both approaches are equally good. Mathematically, the most appropriate way to deal with the delta potential is by using the theory of self-adjoint extension \cite{ajp1,ajp2,gitman}.  The effect of the delta potential is to impose a twisted boundary condition $A_{0^+} = e^{i \phi } A_{2\pi^-} =  e^{i \phi }A_{0^-}$. The question is just what values of $\phi $ will make the momentum operator $-i \partial /\partial k $ a self-adjoint operator \footnote{Although non-self-adjoint operators are interesting too. They should not bother us here. For those interested, see Cuenin J C, Laptev A and Tretter C 2014 Eigenvalue estimates for non-selfadjoint Dirac operators on the real line
Annales Henri Poincar\'e \textbf{15} 707}. The answer is, as can be easily checked, that any real number $\phi$ will do \cite{ajp1,ajp2,gitman}. Therefore, the linking conditions (\ref{link3}) and (\ref{link4}) are just two among the infinitely many possibilities. In a particular problem, whether (\ref{link3}) or (\ref{link4}) or any other value of $\phi $ is realized cannot be determined \emph{a priori}, but can only be inferred from some physical observables (like the spectrum or dynamics of the system). In our specific case,  our previous works \cite{epl1,epl2,prb,ejp19} indicate that it is the linking condition (\ref{link3}) that is realized. This will be verified again in the next section.

In hindsight, it is not difficult to recognize the reason why the spectrum depends only on the end values of $f$. We note that (\ref{eqAk}) is a Dirac-type equation for a charged particle constrained in a ring which is pierced by a magnetic field. The Hamiltonian is
$\mathcal{K} = - i \Delta \partial /\partial k + F_k $, where $k \in [0, 2\pi) $ is the angular variable and $F_k$ is the vector field.
Actually, $H_f$ in (\ref{H}) is just the representation of $\mathcal{K}$ in the basis of the plane waves on the ring, i.e., if we identify level $|n\rangle $ with the basis function $e^{in k }/\sqrt{2\pi }$.

As is well-known and easily checked, (\ref{eqAk}) is invariant under the gauge transform \cite{sakurai}
\begin{eqnarray}
% \nonumber to remove numbering (before each equation)
  A_k &\rightarrow & A_k e^{i \varphi(k)}, \quad F_k \rightarrow F_k + \Delta \frac{\partial \varphi}{\partial k } ,
\end{eqnarray}
where $\varphi(k) $ is an arbitrary, smooth, and $2\pi$-periodic function of $k$.
Under this gauge transform, the value of $f$ at a generic point
\begin{eqnarray}
% \nonumber to remove numbering (before each equation)
  f(n) &=& \frac{1}{2\pi}\int_0^{2\pi } F_k e^{-i  n k } dk
\end{eqnarray}
of course changes. However, the end values $f(0)$ and $f(\infty )$ are invariant, as
\begin{eqnarray}
% \nonumber to remove numbering (before each equation)
  \int_0^{2\pi } \left( \frac{\partial \varphi}{\partial k }\right) e^{-i  n k } dk \bigg|_{n=0} &=& \varphi(2\pi) -\varphi(0)=0,  \\
  \int_0^{2\pi } \left( \frac{\partial \varphi}{\partial k } \right ) e^{-i  n k } dk \bigg|_{n\rightarrow \infty} &=& 0.
\end{eqnarray}
Here the second equation is due to the Riemann-Lebesgue Lemma. Conversely, any two $f$'s with the same end values can be converted into each other with a gauge transform, which leaves the spectrum invariant.

\section{Numerical simulation}

It is instructive to check the analytic results above numerically. To do so, we have to truncate the infinite-dimensional Hilbert space to a finite one. Let us keep the $2N+1$ levels in the middle, i.e., $\{ |-N\rangle, \ldots, |N\rangle \}$. The truncated Hamiltonian is ($\Delta = 1$ below)
\begin{eqnarray}\label{Hfn}
% \nonumber to remove numbering (before each equation)
  H_f^{(N)}  = \sum_{n = -N}^N n \Delta |n \rangle \langle n | +  \sum_{n_1,n_2 = -N}^N f(n_1 - n_2) |n_1 \rangle \langle n_2 | .
\end{eqnarray}
Here for the function $f$,  we shall take (just for the sake of simplicity)
\begin{eqnarray}\label{fnparticular}
% \nonumber to remove numbering (before each equation)
  f(n) &=& g_1 \alpha^{|n|} +g_2 ,
\end{eqnarray}
where $g_{1,2}$ are two real constants and $|\alpha |\leq 1 $ is a decaying factor.

The Hamiltonian $H_f^{(N)}$ can be readily diagonalized numerically and the eigenvalues can be denoted and ordered as $E_{-N}< \ldots < E_0 <  \ldots < E_N $. It is no wonder that the eigenvalues in the middle of the spectrum, which are least affected by the finiteness of the truncated Hilbert space, are indeed almost equally spaced. The only concern is whether the offset agrees with the analytic predictions. We shall therefore focus on the particular eigenvalue $E_0$, which should suffer least from the finite-size effect.

We note that the decaying factor $\alpha $ defines a characteristic length $\lambda= -1/\ln |\alpha|$. Now for $E_0$, there exist two limiting cases. If $N\gg \lambda$, $H_f^{(N)}$ resembles the ideal, infinite-dimensional $H_f$, and it is expected that
\begin{eqnarray}\label{lim1}
% \nonumber to remove numbering (before each equation)
  E_0 &\simeq & g_1 + \frac{1}{\pi}\arctan(\pi g_2),
\end{eqnarray}
according to (\ref{spectrum2}). On the contrary, if $N \ll \lambda $ but still $N\gg1$, all the off-diagonal couplings are close to $g_1 +g_2$, and $H_f^{(N)}$ resembles $H$ in (\ref{H0}) with $g=g_1+g_2$. We thus expect that
\begin{eqnarray}\label{lim2}
% \nonumber to remove numbering (before each equation)
  E_0 &\simeq &  \frac{1}{\pi}\arctan[\pi (g_1+g_2)].
\end{eqnarray}

\begin{figure*}[tb]
\centering
\includegraphics[width= 0.45\textwidth ]{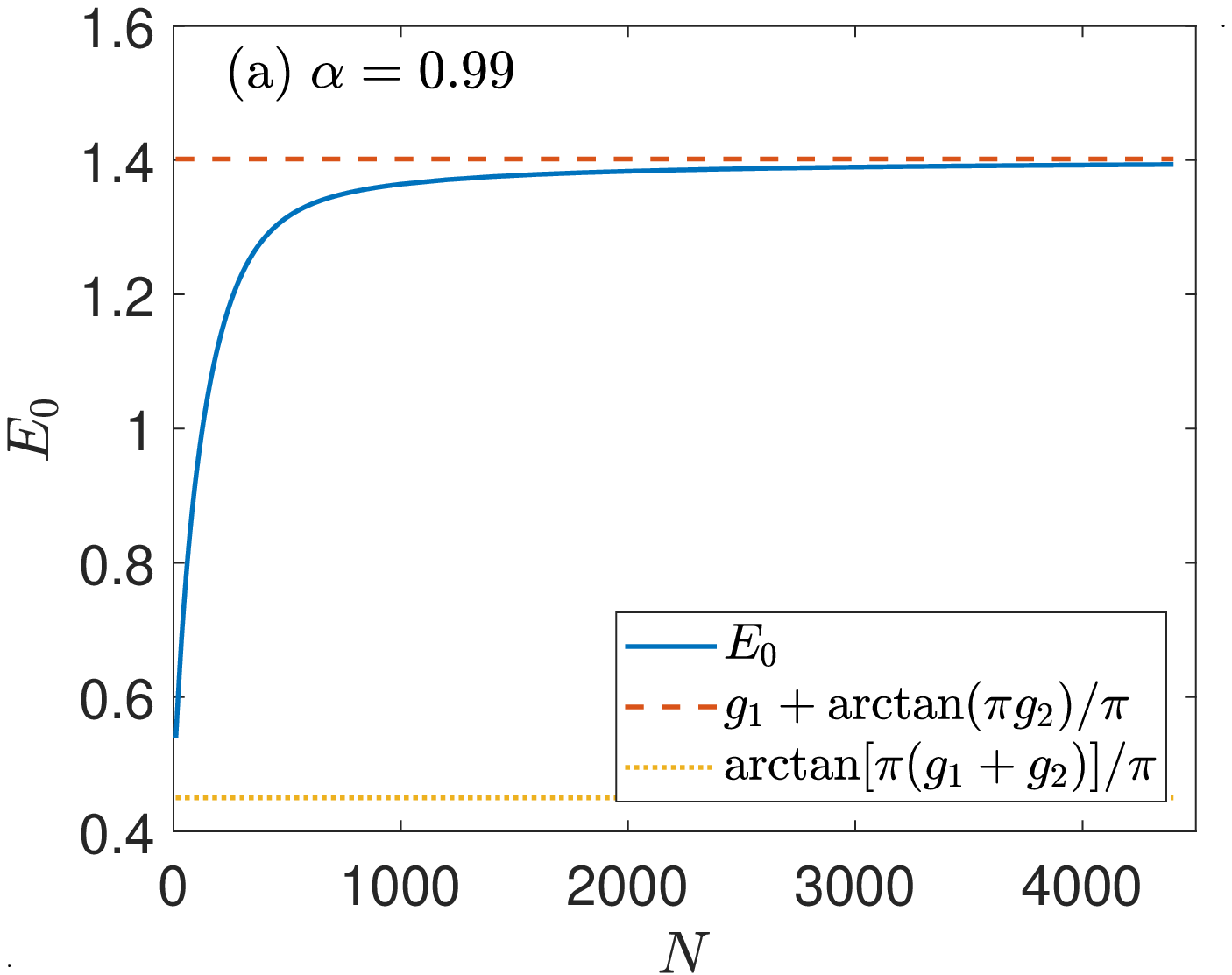}
\includegraphics[width= 0.45\textwidth ]{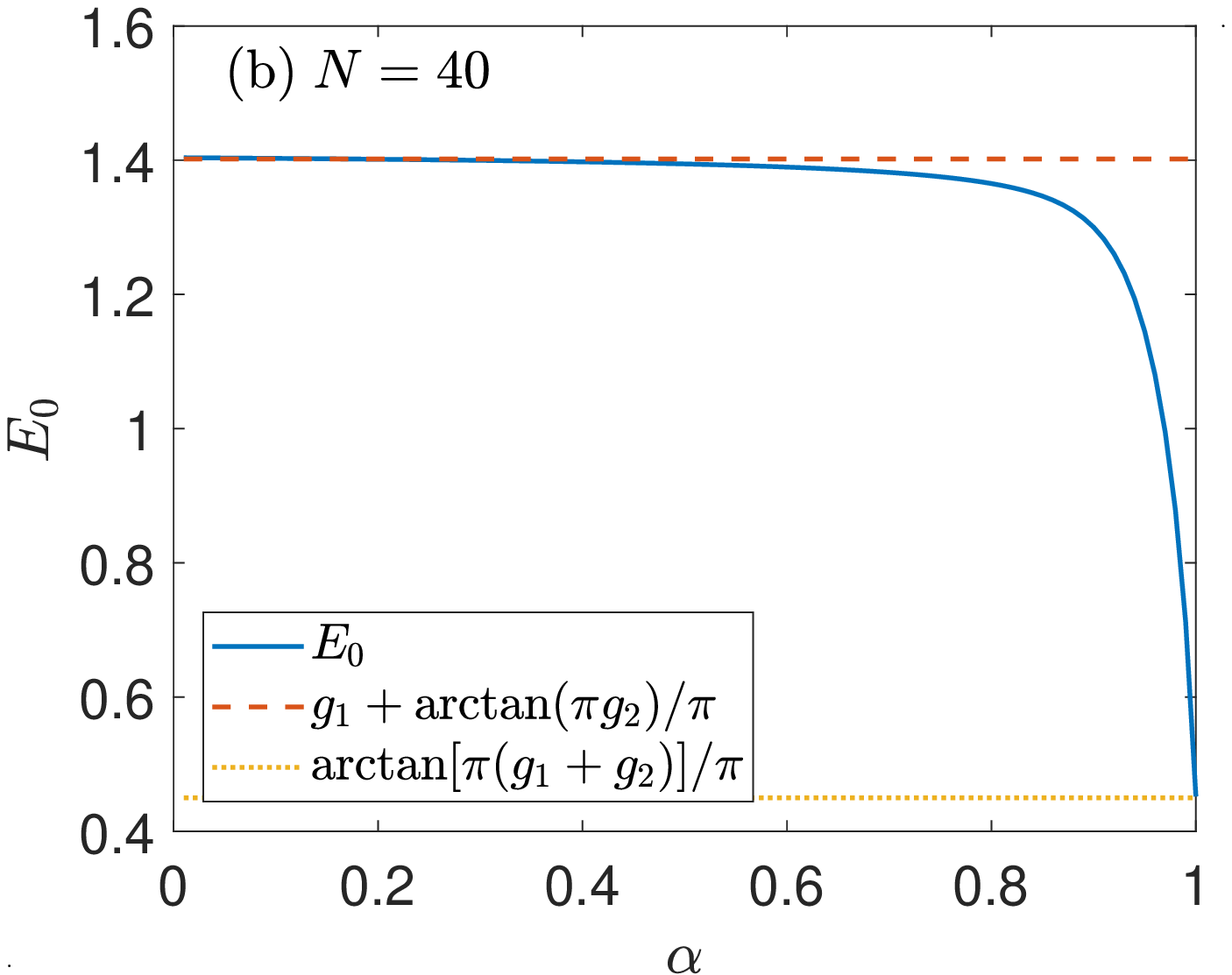}
\caption{(Color online) Crossover behavior of the central eigenvalue $E_0$ of the truncated Hamiltonian $H_f^{(N)}$ as defined in (\ref{Hfn}) and (\ref{fnparticular}). The parameters are $\Delta = g_1 =g_2 =1$. In both panels, the solid line corresponds to the exact value of $E_0$, while the dashed and dotted lines to the analytic predictions in (\ref{lim1}) and (\ref{lim2}), respectively. Note how the solid line approaches the two analytic values at the two ends.   }
\label{fig1}
\end{figure*}

These limiting behaviors are verified in Fig.~\ref{fig1}. In Fig.~\ref{fig1}(a), we fix the value of $\alpha $ to $\alpha =0.99$ (corresponding to a value of $\lambda \simeq 100$) and let $N$ increase from 10 to 4500. We see indeed a crossover between the two values in (\ref{lim1}) and (\ref{lim2}). Similarly, in Fig.~\ref{fig1}(b), we fix $N$ to $N=40$ and let $\alpha$ vary from about $0.01$ to $1$ (in this way, $\lambda $ increases from about 0.2 to infinity). We see the same crossover as in Fig.~\ref{fig1}(a) but in the other direction.

One should have noticed that here the two limits $\lim_{N\rightarrow \infty }$ and $\lim_{\alpha \rightarrow 1^- }$ are not interchangeable. That is,
\begin{eqnarray}
% \nonumber to remove numbering (before each equation)
 g_1 + \frac{1}{\pi}\arctan(\pi g_2) & =& \lim_{\alpha \rightarrow 1^- }  \lim_{N\rightarrow \infty } E_0 \nonumber \\
 &\neq &    \lim_{N\rightarrow \infty } \lim_{\alpha \rightarrow 1^- } E_0 = \frac{1}{\pi}\arctan[\pi (g_1+g_2)].
\end{eqnarray}
This singularity comes essentially from that of the function $\alpha^N $, as we have
\begin{eqnarray}
\lim_{\alpha \rightarrow 1^- }  \lim_{N\rightarrow \infty } \alpha^N = 0 \neq 1 = \lim_{N\rightarrow \infty } \lim_{\alpha \rightarrow 1^- }   \alpha^N.
\end{eqnarray}
 The situation here is somewhat similar to that of a ferromagnet below the Curie point. There, the limit of reducing the external magnetic field to zero and that of increasing the system size to infinity do not commute \cite{huang}. If the former is taken first, the total magnetization would be zero; however, if the latter is taken first, the total magnetization would be nonzero.

\section{Conclusions and discussions}

We have tried to regularize a previously known exactly solvable model which is deemed short of generality. In doing so, we found that it is actually an extremal member of a class of exactly solvable models. It is the Toeplitz property instead of the rank-1 property that has been generalized. Consequently, the solution method is also different.

The interesting thing is that while the model is parameterized by a function $f$, its spectrum depends only on the end values of this function. This characteristic is best understood by noticing that the model can be realized with a charged particle (albeit with the Hamiltonian being linear instead of quadratic in the momentum) confined on a ring which is threaded by some magnetic flux. As long as the end values are fixed, different $f$'s just correspond to different choices of the gauge. But the energy spectrum of the system, as a physical observable, of course should be gauge-independent.
%
%In conclusion, we have solved analytically the eigenvalues and eigenstates of a toy model, which arose in some quantum quench dynamics problem. This on the one hand fills a gap, and on the other hand provides a different (possibly more natural) approach to studying its dynamics.
%
%The toy model should be of some pedagogical value, as the model itself and the solution presented here are all simple enough. It can serve as a good exercise in a undergraduate  course of quantum mechanics.

Although deemed a toy model, we find that the current model contains an interesting model in solid state physics as a special case. Specifically, if $f= -1$ for $x= \pm 1$, and 0 otherwise, the model (\ref{H}) becomes
\begin{eqnarray}
% \nonumber to remove numbering (before each equation)
  H_{Stark} &=&  \sum_{n = -\infty}^\infty n \Delta |n \rangle \langle n | -  \sum_{n = -\infty}^\infty ( |n \rangle \langle n +1 |+|n+1 \rangle \langle n  |) .
\end{eqnarray}
This is just the Hamiltonian of a particle in a tilted one-dimensional tight-binding model. In this system, we have the celebrated Bloch oscillation phenomenon, and the Hamiltonian has been well studied \cite{  glushkov2, blochnjp, glushkov1}. In particular, in \cite{blochnjp}, its spectrum has been solved. It is in agreement with our result (\ref{spectrum1}).

We note that the original model finds use in the quench dynamics of a Bloch state with a potential non-vanishing only on a single site \cite{epl1,epl2,prb}. Hopefully, the current model is relevant if the quench potential is a more general, extended one. But anyway, the model is of pedagogical value for students of quantum mechanics. On one hand, it adds to the short list of exactly solvable models  mentioned in a typical quantum mechanics course (that includes the hydrogen atom, the harmonic oscillator, etc.); on the other hand, the solution is simple enough and illuminating.
An important point is that while the Schr\"odinger equation in the presence of a $\delta$-potential is discussed in possibly every quantum mechanics textbook, the Dirac counterpart is rarely mentioned. As we have seen, the latter is a little bit more tricky.
% Although the toy model is not necessarily of much physical significance, we do believe it is of pedagogical value and can serve as a good exercise in a undergraduate  course of quantum mechanics.

%This work is supported by the National Science Foundation of China under Grant No. 11704070.

\section*{Acknowledgments}

The authors are grateful to K. Jin and Y. Xiang for their helpful comments. This work is supported by the National Science Foundation of China under Grant No. 11704070.


\section*{References}

\begin{thebibliography}{99}

\bibitem{epl1}
Zhang J M  and  Yang H T 2016 Cusps in the quench dynamics of a Bloch state EPL \textbf{114} 60001

\bibitem{epl2}
Zhang J M  and  Yang H T 2016 Sudden jumps and plateaus in the quench dynamics of a Bloch state EPL \textbf{116} 10008

\bibitem{prb}
Zhang J M  and  Liu Y 2018 Dynamical Friedel oscillations of a Fermi sea Phys. Rev. B \textbf{97} 075151

\bibitem{ejp19}
Yang K L and Zhang J M 2019 On an exactly solvable toy model and its dynamics Eur. J. Phys. \textbf{40} 035401

\bibitem{mermin}
Ashcroft N W and Mermin N D 1976 \emph{ Solid State Physics} (Thomson Learning, Toronto)

\bibitem{bocher}
Bochner S and Chandrasekharan K 1949 \emph{Fourier Transforms} (Princeton University Press)

\bibitem{glushkovbook}
Glushkov A V 2008 \emph{Relativistic Quantum Theory: Quantum Mechanics of Atomic Systems} (Astroprint, Odessa)


\bibitem{jpc}
Subramanian R and Bhagwat K V 1972 The relativistic Tamm model J. Phys. C: Solid State Phys. \textbf{5} 798

\bibitem{surf}
Fairbairn W M, Glasser M L and Steslicka M 1973 Relativistic theory of surface states Surf. Sci. \textbf{36} 462

\bibitem{prc}
McKellar B H J and Stephenson G J 1987 Relativistic quarks in one-dimensional periodic structures Phys. Rev. C \textbf{35} 2262

\bibitem{ajp1}
Bonneau G, Faraut J and Valent G 2001 Self-adjoint extensions of operators and the teaching of quantum mechanics Am. J. Phys. \textbf{69} 322

\bibitem{ajp2}
Araujo V S, Coutinho F A B and Perez J F 2004 Operator domains and self-adjoint operators Am. J. Phys. \textbf{72} 203

\bibitem{gitman}
Gitman D M, Tyutin I V and Voronov B L  2012  {\it Self-Adjoint Extensions in Quantum Mechanics} (Springer, Berlin)

\bibitem{sakurai}
Sakurai J J 1994 \emph{Modern Quantum Mechanics} (Addison-Wesley, New York)

\bibitem{huang}
Huang K \emph{Statistical Mechanics}  1987 (Wiley, New York)

\bibitem{glushkov2}
Glushkov A V and Ivanov L N 1993 DC strong-field Stark effect: consistent quantum-mechanical
approach J. Phys. B: At. Mol. Opt. Phys. \textbf{26} L379

\bibitem{blochnjp}
 Hartmann T, Keck F, Korsch H J and Mossmann S 2004 Dynamics of Bloch oscillations New J.
Phys. \textbf{6} 2

\bibitem{glushkov1}
Glushkov A V 2014 Spectroscopy of atom and nucleus in a strong laser field: Stark effect and
multiphoton Resonances J. Phys.: Conf. Ser. \textbf{548} 012020



\end{thebibliography}
\end{document}